\documentclass{elsart}
\usepackage{amssymb}
\usepackage[pctex32]{graphics}
\usepackage{graphicx}
\graphicspath{{converted_graphics/}}

\begin{document}

\begin{frontmatter}

\title{
Noisy-chaotic time series and \\ the forbidden/missing patterns
paradigm}

\author[brazil,calculo,CONICET]{Osvaldo A. Rosso\corauthref{cor}},
\ead{oarosso@fibertel.com.ar}
\corauth[cor]{Corresponding author}
\author[newcastle,brazil]{Laura C. Carpi},
\ead{lauracarpi@gmail.com}
\author[newcastle]{Patricia M. Saco},
\ead{Patricia.Saco@newcastle.edu.au}
\author[brazil2]{Mart\'in G\'omez Ravetti},
\ead{martin.ravetti@dep.ufmg.br}
\author[mdp,CONICET]{Hilda A. Larrondo}, and
\ead{larrondo@fi.mdp.edu.ar}
\author[iflp,gra,palma,CONICET]{Angelo Plastino}.
\ead{plastino@fisica.unlp.edu.ar}

\address[brazil]{Departamento de F\'{\i}sica, Instituto de Ci\^encias Exatas.\\
                 Universidade Federal de Minas Gerais.\\
                 Av. Ant\^onio Carlos, 6627 - Campus Pampulha. \\
                 31270-901 Belo Horizonte - MG, Brazil.}

\address[calculo]{Chaos \& Biology Group,
                  Instituto de C\'alculo, \\
                  Facultad de Ciencias Exactas y Naturales.\\
                  Universidad de Buenos Aires.\\
                  Pabell\'on II, Ciudad Universitaria.\\
                  1428 Ciudad Aut\'onoma de Buenos Aires, Argentina.}

\address[newcastle]{Civil, Surveying and Environmental Engineering.\\
                    The University of Newcastle. \\
                    University Drive, Callaghan NSW 2308, Australia.}

\address[brazil2]{Departamento de Engenharia de Produ\c{c}\~ao, \\
                  Universidade Federal de Minas Gerais. \\
                  Av. Ant\^onio Carlos, 6627, Belo Horizonte, \\
                  31270-901 Belo Horizonte - MG, Brazil.}

\address[mdp]{Facultad de Ingenier\'{\i}a, Universidad Nacional de Mar del Plata.\\
              Av. J.B. Justo 4302, 7600 Mar del Plata, Argentina}

\address[iflp]{Instituto de F\'{\i}sica, IFLP-CCT.\\
             Universidad Nacional de La Plata (UNLP).\\
             C.C. 727, 1900 La Plata, Argentina}

\address[gra]{Instituto Carlos I de F\'{\i}sica Teorica y Computacional and
              Departamento de F\'{\i}sica Atomica, Molecular y Nuclear.\\
              Universidad de Granada, Granada, Spain.}

\address[palma] {Departamento de F\'{\i}sica and IFISC.
                 Universitat de les Illes Balears.\\
                 07122 Palma de Mallorca, Spain.}

\address[CONICET]{Fellow of CONICET-Argentina.}

\begin{abstract}
We deal here with the issue of determinism versus randomness in
time series. One wishes to identify their relative importance
in a given time series. To this end we extend {\it i)\/} the use of ordinal
patterns-based probability distribution functions  associated to a
time series [Bandt and Pompe, Phys. Rev. Lett. 88 (2002) 174102]
and {\it ii)\/} the so-called Amig\'o paradigm of forbidden/missing
patterns [Amig\'o, Zambrano, Sanju\'an, Europhys. Lett. 79 (2007) 50001],
to analyze deterministic finite time series contaminated  with strong additive noises
of different correlation-degree.
Insights pertaining to the deterministic component of the original time series are obtained
with the help of the causal entropy-complexity plane [Rosso {\it et al.\/} Phys. Rev. Lett. 99 (2007) 154102].

PACS: 05.45.Tp; 
      02.50.-r; 
      05.40.-a; 
      05.40.Ca; 
\vskip 0.5 mm
Version: V14
\end{abstract}

\maketitle
\end{frontmatter}

\section{Introduction}
\label{sec:Intro}

\subsection{Preliminaries}
\label{sec:Preliminaries}

The distinction between deterministic and random components in time series
has attracted considerable attention.
From previous research, we may mention here work by  A. R. Osborne and
A. Provenzale \cite{Osborne1989}, G. Sugihara and R. May \cite{Sugihara1990}, D. T. Kaplan and L. Glass
\cite{Kaplan1992,Kaplan1993}. In particular, Kantz and co-workers
\cite{Kantz2000,Cencini2000} analyzed  recently the behavior of
entropy quantifiers as a function of  coarse-graining
resolution, and applied their ideas to the issue of trying to
distinguish between chaos and noise. {\it Why\/} is this an important issue?
Due to the concept of deterministic chaos, derived
from the modern theory of nonlinear dynamical systems, has
profoundly changed our thinking on time-series analysis.
Emphasis is being placed on non-linear approaches, i.e.,
dealing with nonlinear deterministic autonomous equations of
motion representative of chaotic systems that give birth to irregular
signals \cite{Kantz2002}.

Clearly, signals emerging from chaotic time series occupy a
intermediate position  between {\it (a)\/} predictable regular or
quasi-periodic signals and, {\it (b)\/} totally irregular
stochastic signals (noise)  that are completely unpredictable.
However, they exhibit interesting phase-space structures. Chaotic
systems display ``sensitivity to initial conditions" which
 are the origin of  instability everywhere in phase-space. This implies that
instability uncovers information about the phase-space ``population",
not available otherwise \cite{Abarbanel1996}. In turn this leads
us to think of chaos as an {\sl information source,\/} whose
associated rate of generated-information is formulated in precise
fashion via the Kolmogorov-Sinai's entropy
\cite{Kolmogorov1958,Sinai1959}. These considerations motivate our
present interest in the computation of quantifiers based on
Information Theory, like, ``entropy", ``statistical complexity",
``entropy-complexity plane", etc.  These quantifiers
can be used to detect determinism in time series \cite{Rosso2007}.
Indeed, different Information Theory based measures (normalized
Shannon entropy and statistical complexity) allow for a better
distinction between deterministic chaotic and stochastic dynamics
whenever  ``causal" information is incorporated via the Bandt and
Pompe's (BP) methodology \cite{Bandt2002}.

\subsection{A new paradigm: forbidden patterns}
\label{sec:New-paradigm}

When nonlinear dynamics are involved, a deterministic system can generate
``random-looking" results that nevertheless exhibit
persistent trends, cycles (both periodic and non-periodic) and
long-term correlations. Our main interest here, lies in  the
emergence of ``forbidden/missing patterns"
\cite{Amigo2006,Amigo2007,Amigo2008,Amigo2010,Carpi2010}. Why?
Because they  have the potential ability for distinguishing deterministic
behavior (chaos) from randomness in finite time series
contaminated with observational white noise
\cite{Amigo2006,Amigo2007,Carpi2010}. A concomitant helpful
feature, as will be further explained below,  is the decay rate of ``missing ordinal patterns" as a
function of the time series length.

In fact, Zanin \cite{Zanin2008} and Zunino {\it et al.\/}
\cite{Zunino2009} have recently studied the appearance of missing
ordinal patterns in financial time series. The presence of
missing ordinal patterns has also been recently construed as evidence
of deterministic dynamics in epileptic states. Ouyang {\it et al.\/}
\cite{Ouyang2009}  found that a missing patterns' quantifier could
be used as a predictor of epileptic absence-seizures. It is
essential to point out those works have only considered the presence of
uncorrelated noise (white noise), which makes
the associated  results somewhat incomplete, since the  presence
of colored noise might be of importance. A first step in such
direction was recently trodden by us in \cite{Rosso2011}. We
pursue matters here by investigating the robustness of an
associated mathematical construct called the causal
entropy-complexity plane \cite{Rosso2007}, that plays a prominent
role in some of the above cited discoveries. We intend to do this
by analyzing the planes's ability to distinguish between noiseless
chaotic time series and the ones that are contaminated (weekly to
very strongly) with additive correlated noise. The chaotic series
studied here were generated by recourse to a logistic map to which
noise with varying amplitudes was added.

\subsection{Organization of the paper}
Our basic tools are 1) The MPR-statistical complexity
measure together with entropic quantifiers, 2) The Bandt-Pompe
approach to extract causal probability distributions functions
(PDFs) from a given time-series, and 3) the construction of an
entropy-complexity plane. These subjects are covered in great
detail in \cite{Carpi2010,Rosso2011,Rosso2010A}.
For completeness, some details are given in the Appendix.
The reader unfamiliar with these themes is strongly advised to go to
the Appendix at this point.
The forthcoming Section \ref{sec:Logistica} details the problem
to be discussed, while Section \ref{sec:Amigo-paradigm} revisits
the Amig\'o paradigm.
Our results are presented and discussed in Section \ref{sec:Results},
while some conclusions are drawn in Section \ref{sec:Conclusions}.
Background materials are provided in the Appendix.

\section{Logistic map plus observational noise}
\label{sec:Logistica}

The logistic map constitutes  a canonic example, often employed to
illustrate new concepts and/or methods for the analysis of
dynamical systems. Here we will use the logistic map with additive
correlated noise in order to exemplify the behavior of the
normalized Shannon entropy ${\mathcal H}_S$ and the MPR-complexity
${\mathcal C}_{JS}$, both evaluated using a PDF based  on the
Bandt-Pompe's procedure. We will also investigate the behavior of
``missing ordinal patterns" in both {\it a)\/} the logistic map
with additive noise (observational noise) and {\it b)\/} a
pure-noise series, where the number of time series data was fixed at $N$.

The logistic map is a polynomial  mapping of
degree 2, $F: x_n \rightarrow x_{n+1}$ \cite{Sprott2004}, described by
the ecologically motivated,
dissipative system represented by the first-order difference equation
\begin{equation}
x_{n+1} =  r \cdot x_n \cdot ( 1 - x_n )  \ ,
\label{logistica}
\end{equation}
with $0 \leq x_n \leq 1$ and $0 \leq r \leq 4$.

Let $\eta^{(k)}$ be a correlated noise with $f^{-k}$ power spectra
generated as described in \cite{Rosso2007,Rosso2011}.
The steps to be followed are enumerated below.
\begin{enumerate}
\item Using the Mersenne twister generator \cite{Mersenne1998} through
the $\textsc{Matlab}^\copyright$ {\sl rand\/} function we generate
pseudo random numbers $y^0_i$ in the interval $(-0.5,0.5)$ with an
{\it (a)\/} almost flat power spectra (PS),
{\it (b)\/} uniform PDF, and
{\it (c)\/} zero mean value.
\item The Fast Fourier Transform (FFT) ${y^1_i}$ is first obtained
and then multiplied by $f^{-k/2}$, yielding ${y^2_i}$. Then,
${y^2_i}$ is symmetrized so as to obtain a real function.
Subsequently the pertinent inverse FFT is found, after discarding the
small imaginary components produced by the numerical
approximations. The resulting noisy time series is then re-scaled
to the interval $[-1,1]$, which produces a new  time series
$\eta^{(k)}$ that exhibits the desired power spectra and, by
construction,  is representative of non-Gaussian noises.
\end{enumerate}

We consider time series of the form $\mathcal S = \{ S_n, n = 1, \cdots, N \}$
generated  by the discrete system:
\begin{equation}
S_n = x_n + A \cdot \eta^{(k)}_n \ ,
\label{series}
\end{equation}
in which, $x_n$ is given by the logistic map and  $\eta^{(k)}_n \in [-1,1]$ represents
a noise with power spectrum  $f^{-k}$ and amplitude $A$.

In generating the logistic map's component of our time-series we fix $r=4$ and  start
the iteration procedure with a random initial condition.
The first $5\cdot10^4$ iterations are considered part of the transient behavior and discarded.
After the transient part dies out, $N=10^5$ values are generated.
In the case of the stochastic component of our time-series we consider $0 \leq k \leq 2$
with $k$-values changing by an amount $\Delta k=1$
(other intermediate values have been considered in \cite{Rosso2011}).
The noise amplitude was varied in order to cover different regimes from weak
(the noise can be consider as perturbation to the logistic) to
very strong (the logistic is  consider a perturbation to the noise).
Ten noisy time-series of length $N=10^5$ values and using different seeds, are generated for each
value of the $k$-exponent.

\section{The Amig\'o-paradigm}
\label{sec:Amigo-paradigm}

We arrive at the central point of our discussion. For
deterministic one dimensional maps, Amig\'o {\it et al.\/}
\cite{Amigo2006,Amigo2007,Amigo2008,Amigo2010} have conclusively
shown that not all the possible ordinal patterns (as defined using
Bandt and Pompe's methodology, see Appendix) can be effectively
materialized into orbits, which in a sense makes these patterns
``forbidden". We insist: {\it this is an established fact, not a
conjecture}. The existence of these {\it forbidden ordinal
patterns\/} becomes a persistent feature, a ``new" dynamical
property. For a fixed pattern-length (embedding dimension $D$) the
number of forbidden patterns of a time series (unobserved
patterns) is independent of the series length $N$. It must be noted, that
this independence does not characterize other properties of the
series such as proximity and correlation, which die out with time
\cite{Amigo2007,Amigo2010}. For example, in the time series
generated by the logistic map $x_{k+1} = 4 x_k (1 - x_k)$, if we
consider patterns of length $D=3$, the pattern $\{2,1,0\}$ is
forbidden. That is, the pattern $x_{k+2} < x_{k+1} < x_{k}$ never
appears \cite{Amigo2007}.

Stochastic processes could also have forbidden patterns \cite{Rosso2011}.
However, in the case of uncorrelated (white noise) or  certain correlated stochastic processes
(noise with power low spectrum $f^{-k}$ with $k \geq 0$, ordinal
Brownian motion, fractional Brownian motion, and fractional Gaussian noise),
it can be  numerically shown that  {\it no\/}  forbidden patterns emerge.
In the case of time series generated  by an {\it unconstrained stochastic process\/}
(uncorrelated process) every ordinal pattern has the same probability of appearance
\cite{Amigo2006,Amigo2007,Amigo2008,Amigo2010}.
If the time series is long  enough, all the ordinal patterns should eventually appear.
If the number of time-series' observations is sufficiently big, the associated probability
distribution function should be the uniform distribution, and the number of observed patterns
should depend only on the length $N$ of the time series under study.

For correlated stochastic  processes the probability of observing individual patterns depends
not only on the time series length $N$ but also on the correlation structure \cite{Carpi2010}.
The existence of a non-observed ordinal pattern does not qualify it as ``forbidden'', only as
{\it ``missing''\/}, and is due to the finite length of the time series.
A similar observation also holds for the case of real data series, as they always possess a stochastic
component due to the omnipresence of dynamical noise \cite{Wold1938,Kurths1987,Cabanis1988}.
The existence of ``missing ordinal patterns" could be either related to stochastic processes
(correlated or uncorrelated) or to deterministic noisy processes, which is the case for
observational time series.

\subsection{The Carpi-Amig\'o test}
\label{sec:Carpi-test}

Amig\'o and co-workers \cite{Amigo2006,Amigo2007}  proposed a test that uses missing ordinal patterns
to distinguish determinism (chaos) from pure randomness in finite time series contaminated with
observational white noise (uncorrelated noise).
The concomitant methodology \cite{Amigo2007} involves a graphic comparison between:
\begin{itemize}
\item The decay rate of the missing ordinal patterns (of length $D$) of the time series under analysis as a
function of the series length $N$, and
\item the decay rate exhibited by white Gaussian noise.
\end{itemize}
This methodology was recently extended by Carpi {\it et al.\/}
\cite{Carpi2010} for the analysis of missing ordinal patterns in
stochastic processes with different degrees of correlation. We are
speaking of  fractional Brownian motion (fBm), fractional Gaussian
noise (fGn), and noises with $f^{-k}$ power spectrum (PS) and $k \geq 0$.
Results show that for a fixed pattern length, the decay rate
of missing ordinal patterns in stochastic processes depends not
only on the series length but also on their correlation
structures. In other words, missing ordinal patterns are more
persistent in the time series with higher correlation structures.
Carpi {\it et al.\/} \cite{Carpi2010} have also  shown that the
standard deviation of the estimated decay rate of missing ordinal
patterns ($\alpha$) decreases with increasing $D$. This is due to
the fact that longer patterns contain more temporal information
and are therefore more effective in capturing the dynamic of time
series with correlation structures.

An important quantity for us,  called ${\mathcal M}(N,D)$, is the
number of missing ordinal patterns of length $D$ {\it not\/}
observed in a time series  with $N$ values. As we mentioned
before, for pure correlated stochastic processes the probability
of observing an individual pattern of length $D$ depends on the
time series-length $N$ and on the correlation structure (as
determined by  the type of noise $k>0$ ). In fact, as established
 for noises with an $f^{-k}$ PS in \cite{Carpi2010}, as the value
of $k>0$ augments -- which implies that correlations grow --
increasing values of $N$ are needed in reaching the ``ideal"
condition ${\mathcal M}(N,D)=0$. If the time series is chaotic but
has an additive stochastic component, then one expects that as the
time series' length $N$ increases, the number of ``missing ordinal
patterns" will decrease and eventually vanish. That this may
happen does not only depend on  the length $N$
and on the underlying deterministic components of the time series
but also on the correlation-structure of the added noise.

\section{Present results}
\label{sec:Results}

We wish to analyze and characterize the behavior of  finite, noisy
and chaotic time-series by recourse to patterns generated in the
(causal) entropy-complexity plane. We intend to assess in
particular the planar-geography of the forbidden patterns.
The focus of attention is centered upon the roles of {\it i)\/} the noise
amplitude $A$ and {\it ii)\/} the type of contaminating noise
(degree of correlation). We consider noises with $f^{-k}$-power
spectrum.

For our present analysis, we fix the pattern-length at $D=6$, the
embedding time lag at $\tau = 1$, and the time series length at
$N=10^5$ values. For each one of the ten time series series
generated (see Eq. (\ref{series})) and for each pair $(A,k)$, the
normalized Shannon entropy ${\mathcal H}_S$ and the
MPR-statistical complexity ${\mathcal C}_{JS}$ were evaluated
using the  Bandt and Pompe PDFs. We deal with additive
(observational) noise  with $k= 0, 1,$ and $2$, and analyze their
position (each point results from an average over 10 different
series) in the ${\mathcal H} \times {\mathcal C}$-planar graphs. A
previous study has extensively dealt with the subject, but only in
a special  scenario in which the noise is of perturbative
character \cite{Rosso2011}. Here we deal with the behavior of
$(A,k)$ in the planar-graphs way beyond  the perturbative
$A$-zone, by considering also $A>1$ values.

\subsection{Case $0 \leq A \leq 1$}
\label{sec:Amenorque1}

\begin{figure}
\noindent
\includegraphics[width=5.5in]{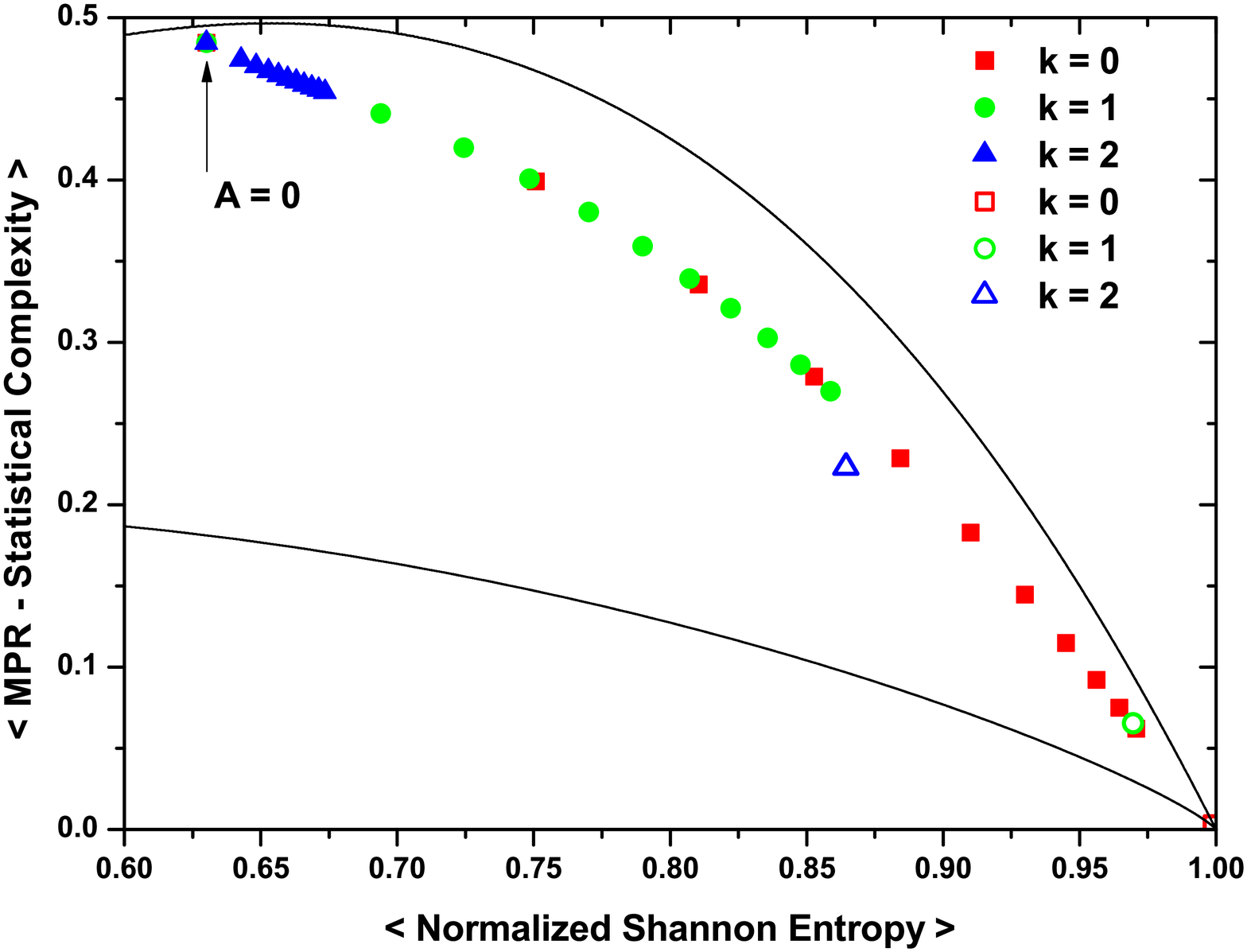}
\caption{
The causality entropy-complexity plane for all time series values (mean) corresponding to
noise amplitude $0\leq A \leq 1$ ($\Delta A=0.1$) and $k=0,1,2$.
Values for strictly noisy time series  are also shown as open symbols.
The  values shown were obtained from times series of length $N=10^5$, 
and Bandt-Pompe parameters $D=6$, $\tau=1$.
The lines represent the values of minimum and maximum statistical complexity ${\mathcal C}_{min}$ 
and ${\mathcal C}_{max}$ evaluated for the case of pattern length $D=6$.
}
\label{plano-k=123}
\end{figure}

Figure \ref{plano-k=123} summarizes the planar behavior of the
noisy chaotic time series considered (the logistic map with
additive correlated noise with $f^{-k}$ PS), considering noise amplitudes
$0 \leq A \leq 1$ ($\Delta A = 0.1$) and $k=0$ (uncorrelated),
$k=1,2$ (correlated), see also Fig. 8 in \cite{Rosso2011}.
Clearly, given  the noise-amplitude values
here considered, the noise has a mere perturbative role vis-a-vis
of the logistic map's chaotic behavior.

For $A=0$ we obtain the purely deterministic value corresponding
to the logistic time-series, localized approximately at a
medium-high entropic coordinate, near the highest possible
complexity value. Note that such is the typical behavior observed for
deterministic systems \cite{Rosso2007}. For a purely uncorrelated
stochastic process ($k=0$) we have ${\mathcal H}_S = 1$ and
${\mathcal C}_{JS} = 0$. The correlated (colored) stochastic
processes ($k \neq 0$) yield points located at intermediate values
between the curves ${\mathcal C}_{min}$ and ${\mathcal C}_{max}$,
with decreasing values of entropy and increasing values of
complexity as $k$ grows \cite{Rosso2007} (see Fig.
\ref{plano-k=123}, open symbols). From this figure and, also from
our previous work \cite{Rosso2011}, we see that if $k \approx 0$,
for increasing values of the amplitude $A$, entropy and
complexity values change (starting from the value corresponding to
the pure logistic series, i.e., $A=0$) with a tendency to approach
the values corresponding to pure noise, that is, (${\mathcal H}_S
\approx 1$ and ${\mathcal C}_{JS} \approx 0$).
Similar behavior is observed when correlated noises are
considered, $k \neq 0$.

The main effect of the additive noise ($A
\neq 0$) is to shift  the point representative of  zero noise
($A=0$) towards increasing values of entropy ${\mathcal H}_S$ and
decreasing values of complexity ${\mathcal C}_{JS}$.
This shift defines a kind of ``trajectory" (curve) in the
${\mathcal H} \times {\mathcal C}$-plane,
that is located in the vicinity of the maximum
complexity ${\mathcal C}_{max}-$curve. Moreover, we found (by
looking at  other chaotic maps) that this trajectory is characteristic
of the dynamical system under analysis.
Such behavior can be linked to the persistence of forbidden patterns because
they, in turn, imply that the deterministic logistic component is still
operative and  influencing the time series' behavior
\cite{Rosso2011}. As $k$ increases, the dependence on the
noise-amplitude $A$ ($0 \leq A \leq 1$) tends to become
attenuated, implying that ${\mathcal M}(N,D) \neq 0$ (see Fig. 4
of \cite{Rosso2011}), while it almost disappears for $k \approx
2$. This fact is due to the effect of the ``coloring" correlations
of the noise, that grow with increasing values of $k$.

\subsection{Case $A \geq 1$}
\label{sec:Amayorque1}

Here we are interested in the physics associated to $A$'s growth,
entailing  an interchange between the  roles of the logistic map
and of the noise because a large $A$ makes noise the dominant
feature. Several effects should be considered
\begin{itemize}
\item noise contamination level (represented by $A$),
\item noise correlation (represented by $k$) and,
\item persistence of the forbidden patterns.
\end{itemize}
We should expect that as $A$ grows (with fixed
values for $N$ and $D$), the  missing-patterns numbers would diminish and
eventually vanish. With strong or very strong noise we expect zero
missing patterns and eventual predominance of stochastic features.
This means ${\mathcal M}(N,D)=0$ independently of the $A$ value.
However, taking into account our previous results
\cite{Rosso2011}, we expect that some new specific features will
be revealed by the use of the causality entropy-complexity plane.

\begin{figure}
\noindent
\includegraphics[width=5in]{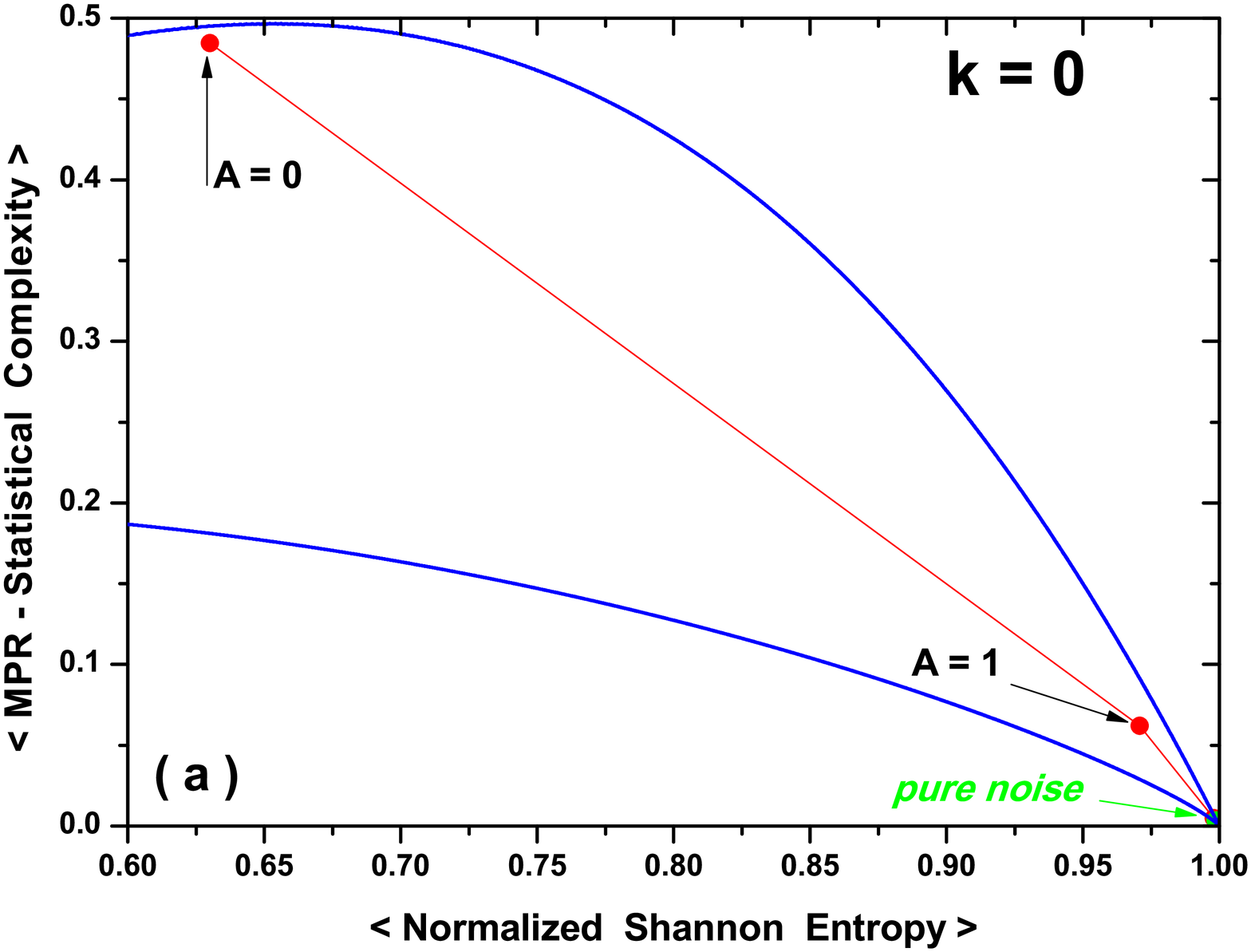}
\includegraphics[width=5in]{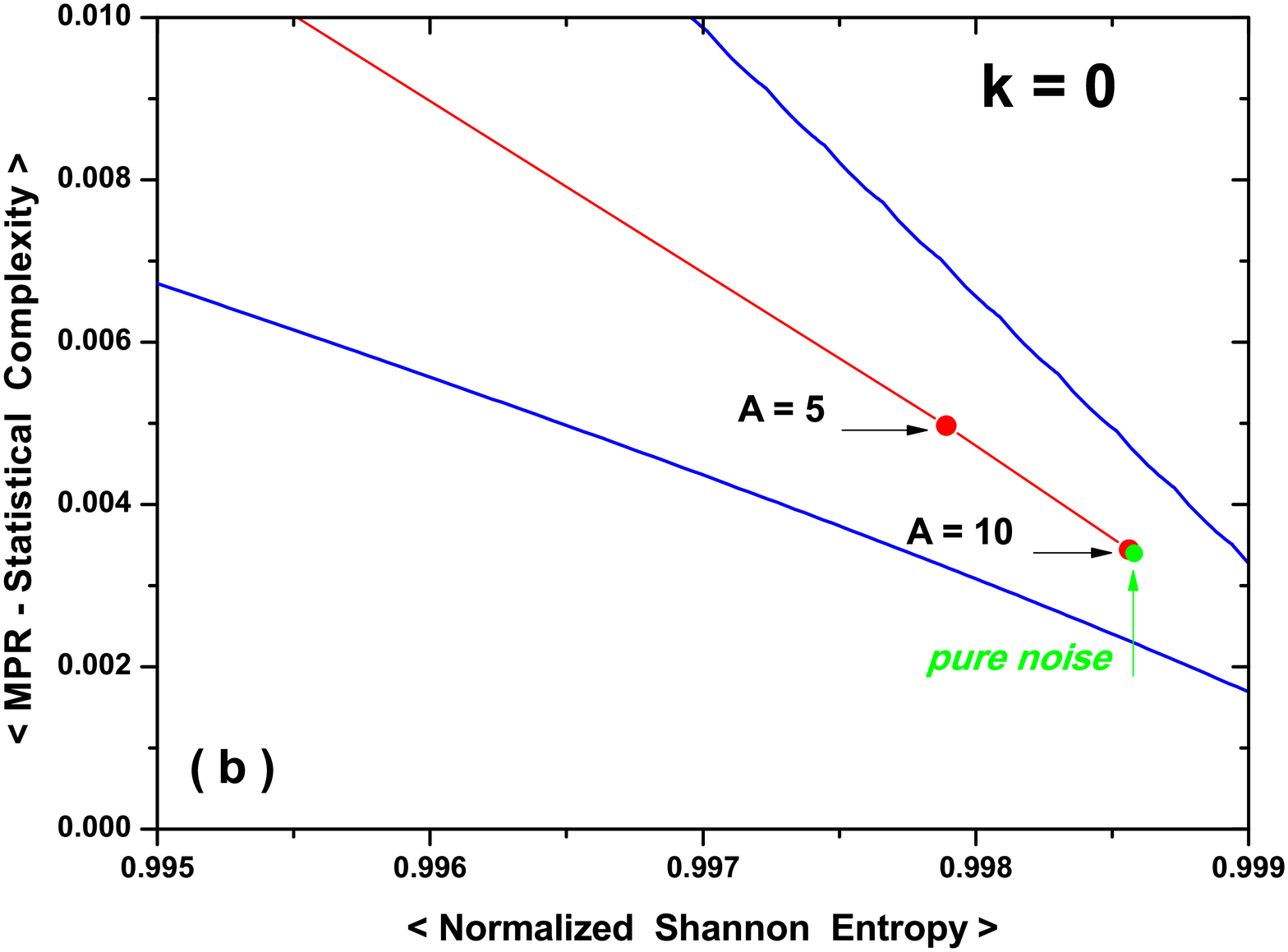}
\caption{
{\it a )\/} The causality entropy-complexity plane for all time series values (mean) corresponding to increasing values of
noise amplitude $A$ and $k=0$.
Value for the case of strictly noisy time series is shown as open symbol.
The  values shown were obtained from times series of length $N=10^5$, 
and Bandt-Pompe parameters $D=6$, $\tau=1$.
The lines represent the values of minimum and maximum statistical complexity ${\mathcal C}_{min}$ 
and ${\mathcal C}_{max}$ evaluated for the case of pattern length $D=6$.
{\it b)\/} Detail for high entropy values.
}
\label{plano-k=0}
\end{figure}

\begin{figure}
\noindent
\includegraphics[width=5in]{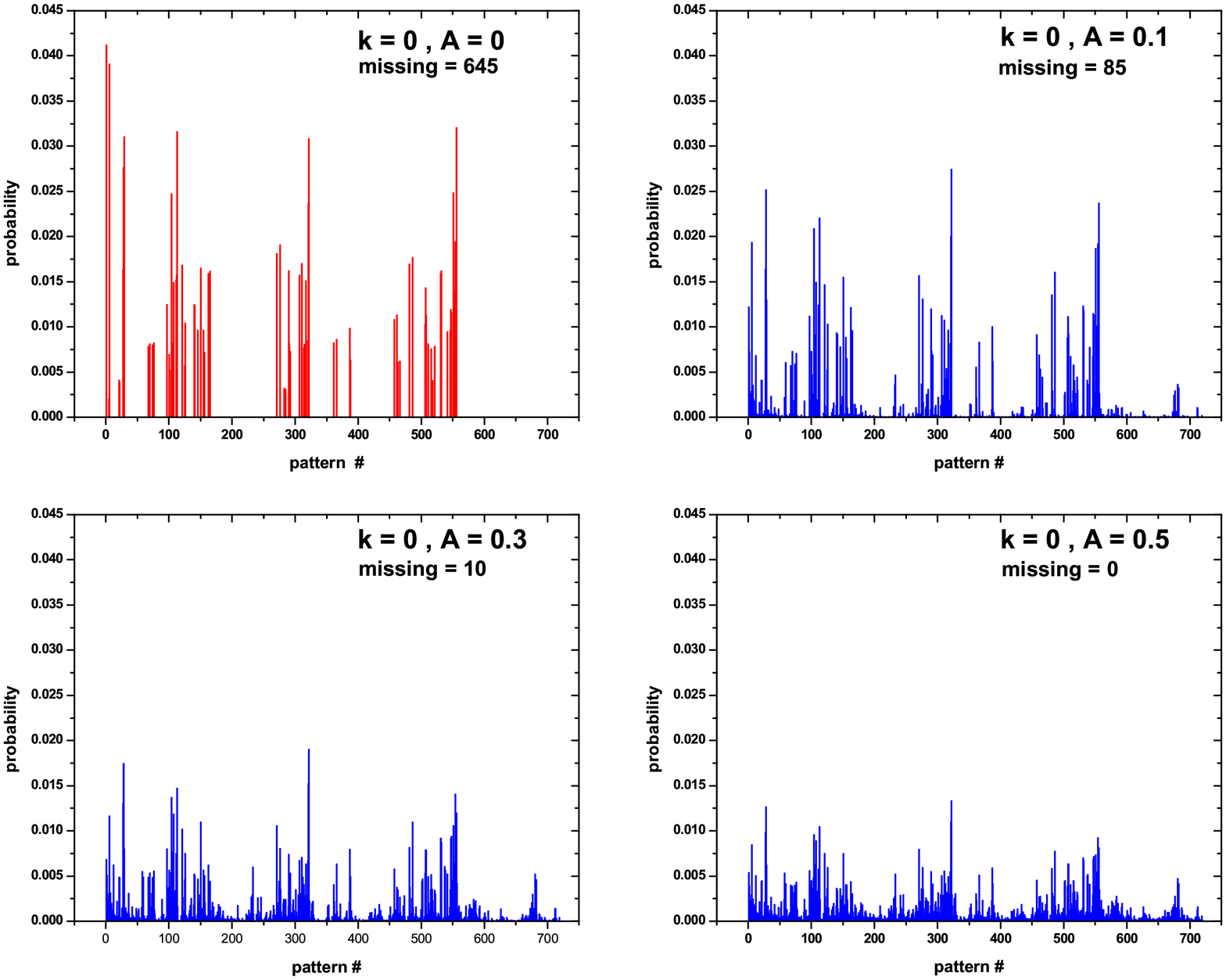}
\includegraphics[width=5in]{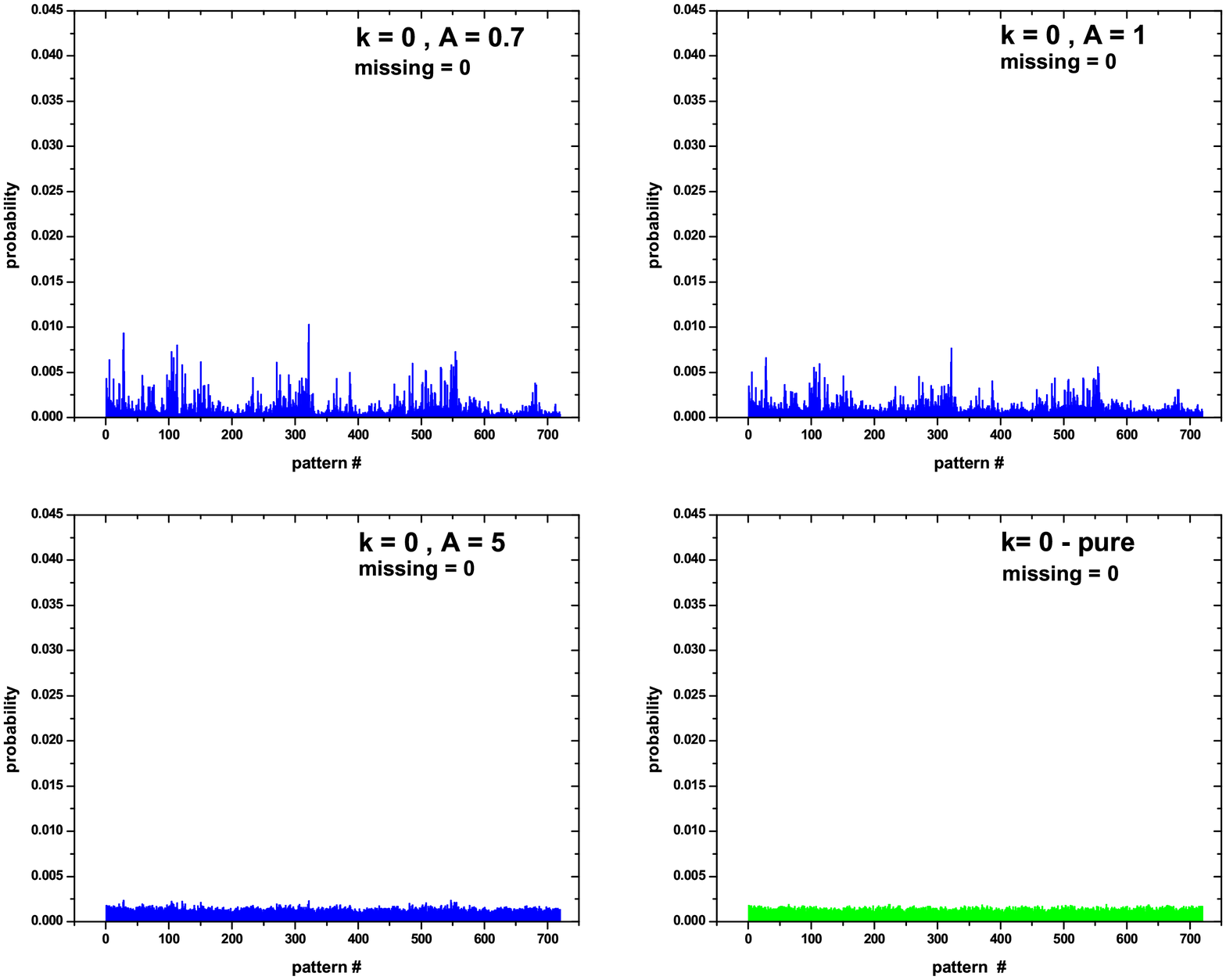}
\caption{
The Bandt-Pompe PDF ($M=720$) for a typical noisy chaotic time series considered (uncorrelated noise $k=0$),
for different increasing values of the noise amplitude $A$.
In the top-left corner the PDF corresponding to the unperturbed logistic map ($A=0$) is presented.
In the bottom-right corner, the PDF for a pure noise $k=0$ time series is display.
The  values shown were obtained from times series of length $N=10^5$, 
and Bandt-Pompe parameters $D=6$, $\tau=1$.
}
\label{PDF-k=0}
\end{figure}

Consider our results for uncorrelated noise $k=0$ and increasing
values of noise-amplitude $A$ depicted in Fig. \ref{plano-k=0}.
This graph reveals that for $A \geq 1$  the points corresponding to increasing values of $A$
continue to move in the plane towards the site representative  of
pure noise, with ${\mathcal M}(N,D)=0$. The perturbative character
of the noise is, of course, lost. This behavior is apparent in
Fig. \ref{PDF-k=0}, where the corresponding Bandt-Pompe PDF's, for
a typical noisy chaotic time-series, are displayed for increasing
noise amplitudes $A$. The corresponding value of
${\mathcal M}(N,D)$ can also be observed in these graphs.

At the top-left corner of Fig. \ref{PDF-k=0} we depict the Bandt-Pompe's PDF
for the unperturbed logistic map ($A=0$).
The presence of forbidden patterns (that have probabilities $p_i =0$) is clearly visible.
In the same plot, at the bottom-right corner, we display the pure noise $k=0$
(white noise) PDF.
No forbidden patterns  exist now, and we observe the characteristic flat
distribution, $p_i \cong 1/M$ for all patterns $i=1, \cdots , M$ ($M=D!=720)$.
The noise-influence on the  PDFs is clearly displayed in this graph (see also the plots
for $k=1$ and $k=2$, Figures \ref{PDF-k=1} and \ref{PDF-k=2}).
One appreciates the fact that the PDF evolves from that of the logistic map towards
that of the pure-noise as one increases the value of  $A$.
The noise-effect also destroys the forbidden character of some patterns, whose
associated probabilities grow slowly with the control parameter $A$, a feature that emphasizes
the presence of a deterministic component
in the noisy time series, even for ${\mathcal M}(N,D) =0$, if $A$ is not too large.
This effect disappears for very strong  noise.
Note that for $A=5$ the PDF is practically equal to that of pure noise. The two PDFs
are almost indistinguishable at $A=10$, and their  planar
localizations coincide (see Fig. \ref{plano-k=0}).

\begin{figure}
\noindent
\includegraphics[width=5in]{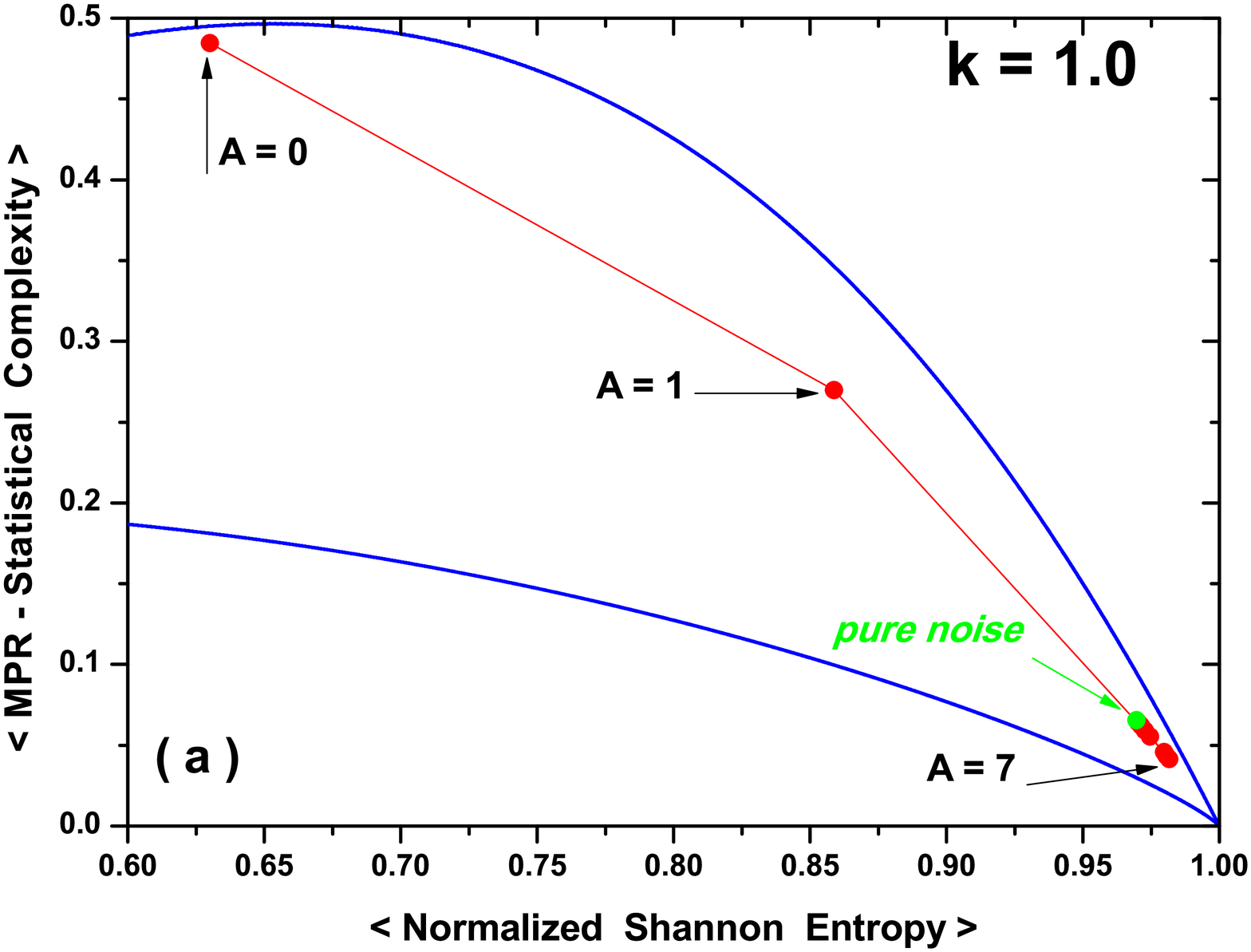}
\includegraphics[width=5in]{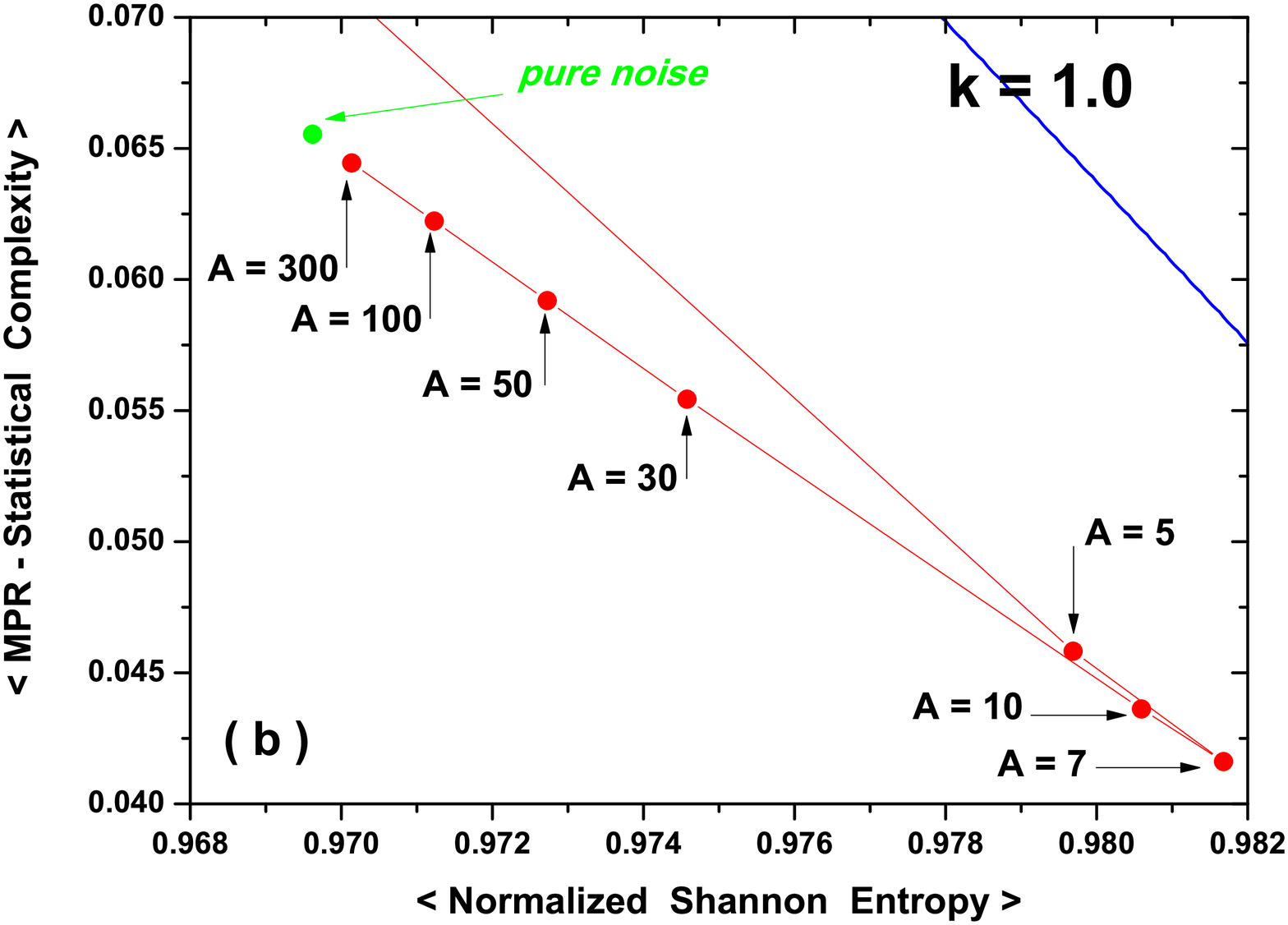}
\caption{
{\it a)\/} The causality entropy-complexity plane for all time series values (mean) corresponding to increasing values of
noise amplitude $A$ and $k=1$.
Value for the case of strictly noisy time series is shown as open symbol.
The  values shown were obtained from times series of length $N=10^5$, 
and Bandt-Pompe parameters $D=6$, $\tau=1$.
The lines represent the values of minimum and maximum statistical complexity ${\mathcal C}_{min}$ 
and ${\mathcal C}_{max}$ evaluated for the case of pattern length $D=6$.
{\it b)\/} Detail for high entropy values.
}
\label{plano-k=1}
\end{figure}

\begin{figure}
\noindent
\includegraphics[width=5in]{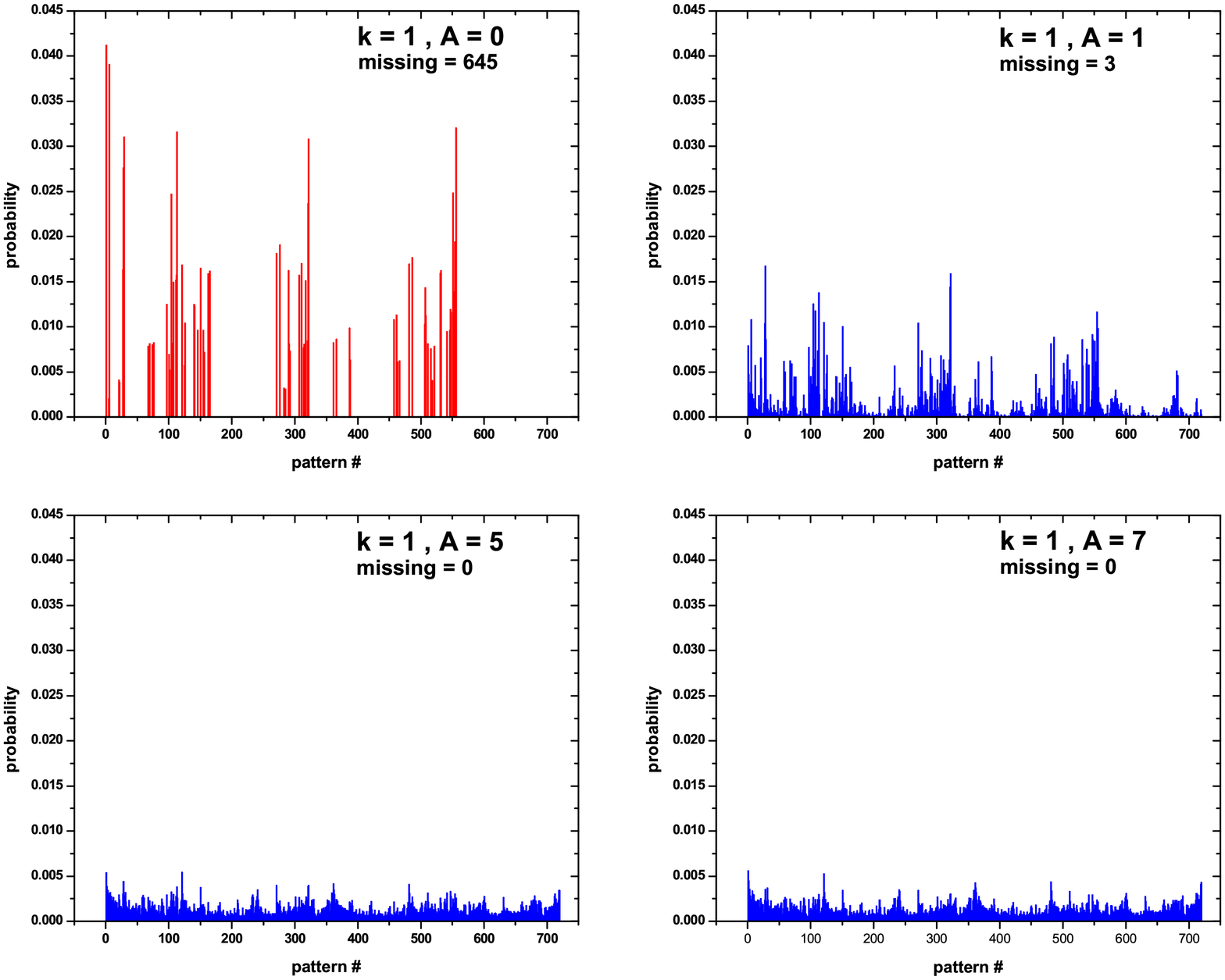}
\includegraphics[width=5in]{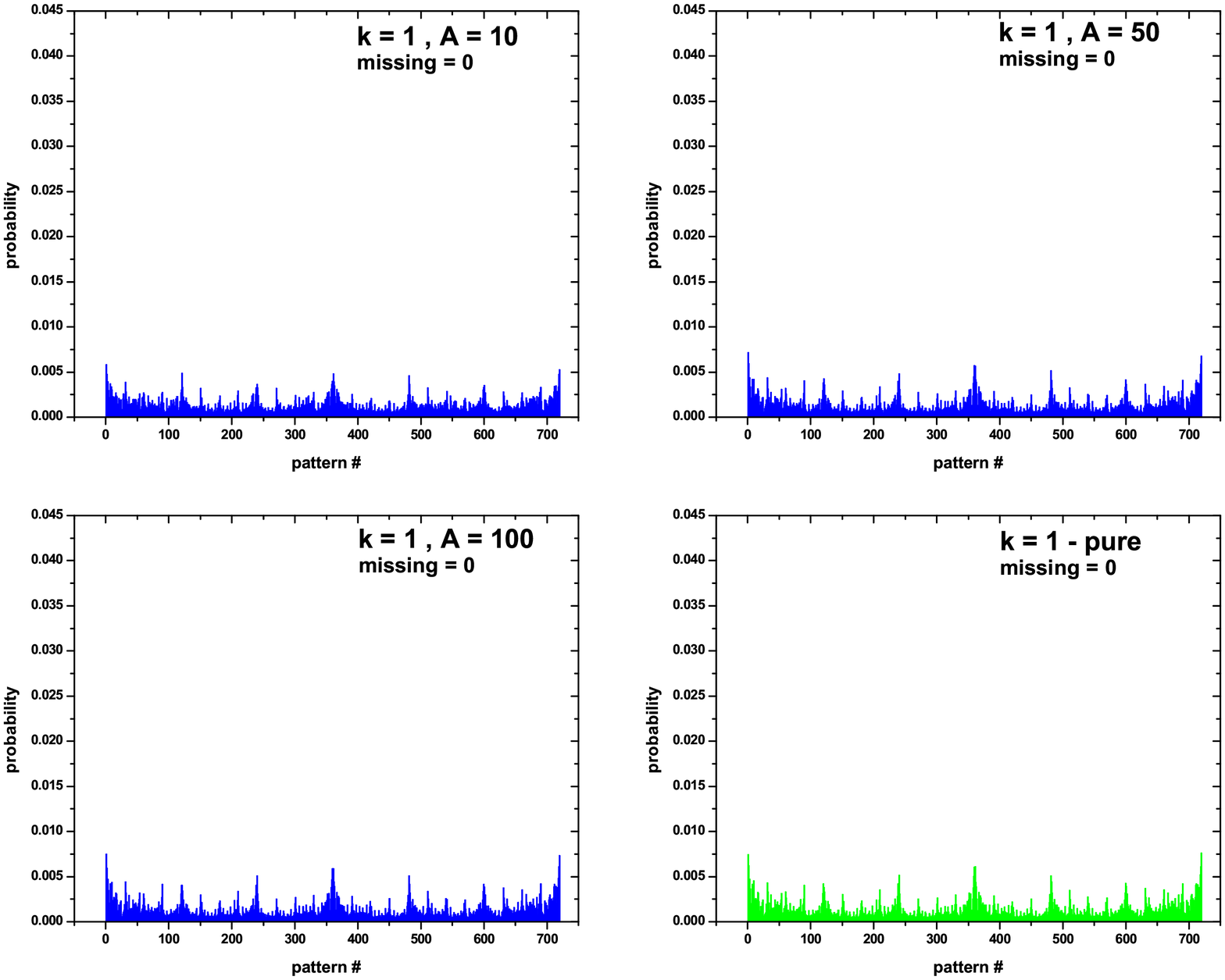}
\caption{
The Bandt-Pompe PDF ($M=720$) for a typical noisy chaotic time series considered (correlated noise $k=1$),
for different increasing increasing values of the noise amplitude $A$.
In the top-left corner the PDF corresponding to the unperturbed logistic map ($A=0$) is presented.
In the bottom-right corner, the PDF for a pure noise $k=1$ time series is display.
The  values shown were obtained from times series of length $N=10^5$, 
and Bandt-Pompe parameters $D=6$, $\tau=1$.
}
\label{PDF-k=1}
\end{figure}

\begin{figure}
\noindent
\includegraphics[width=5in]{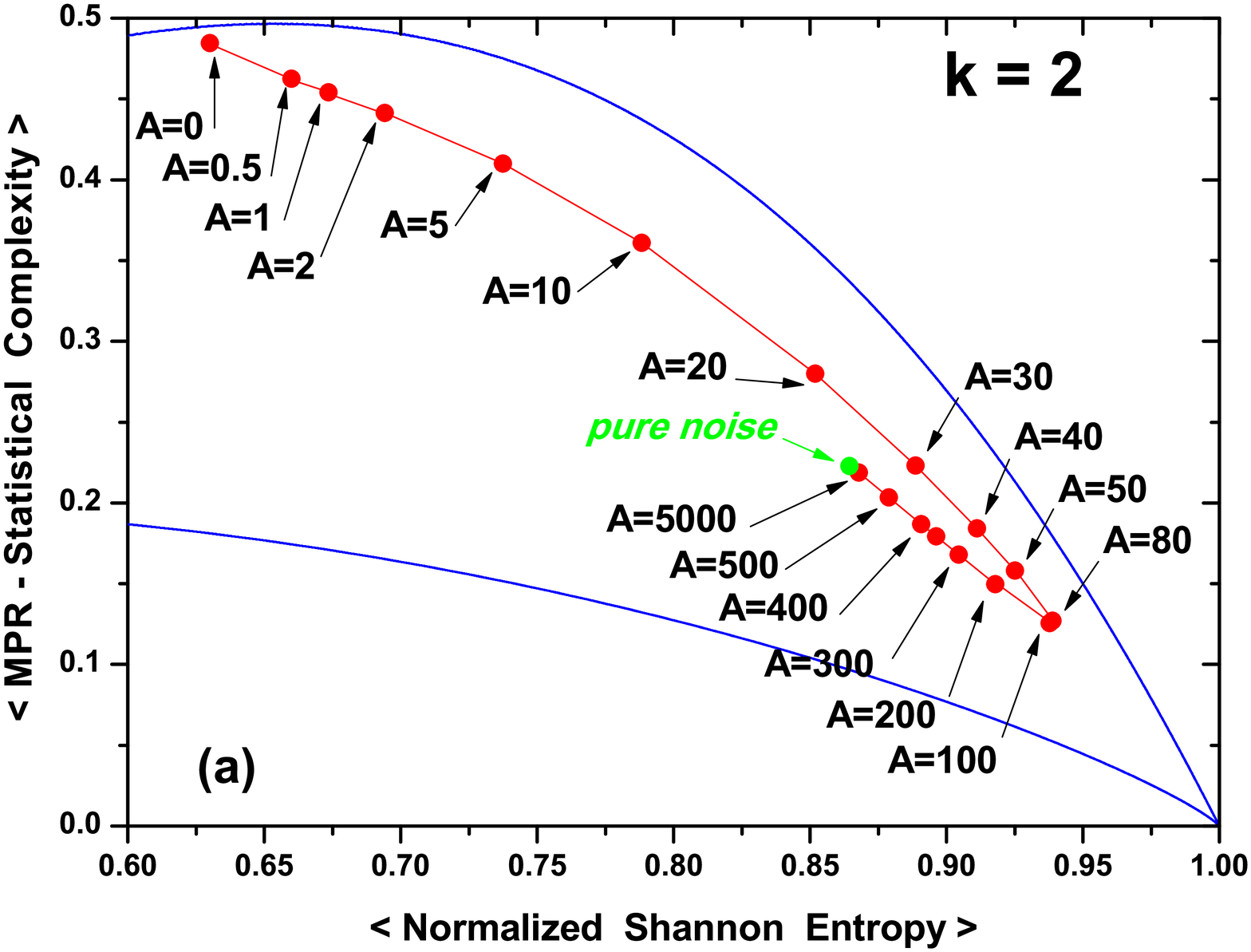}
\includegraphics[width=5in]{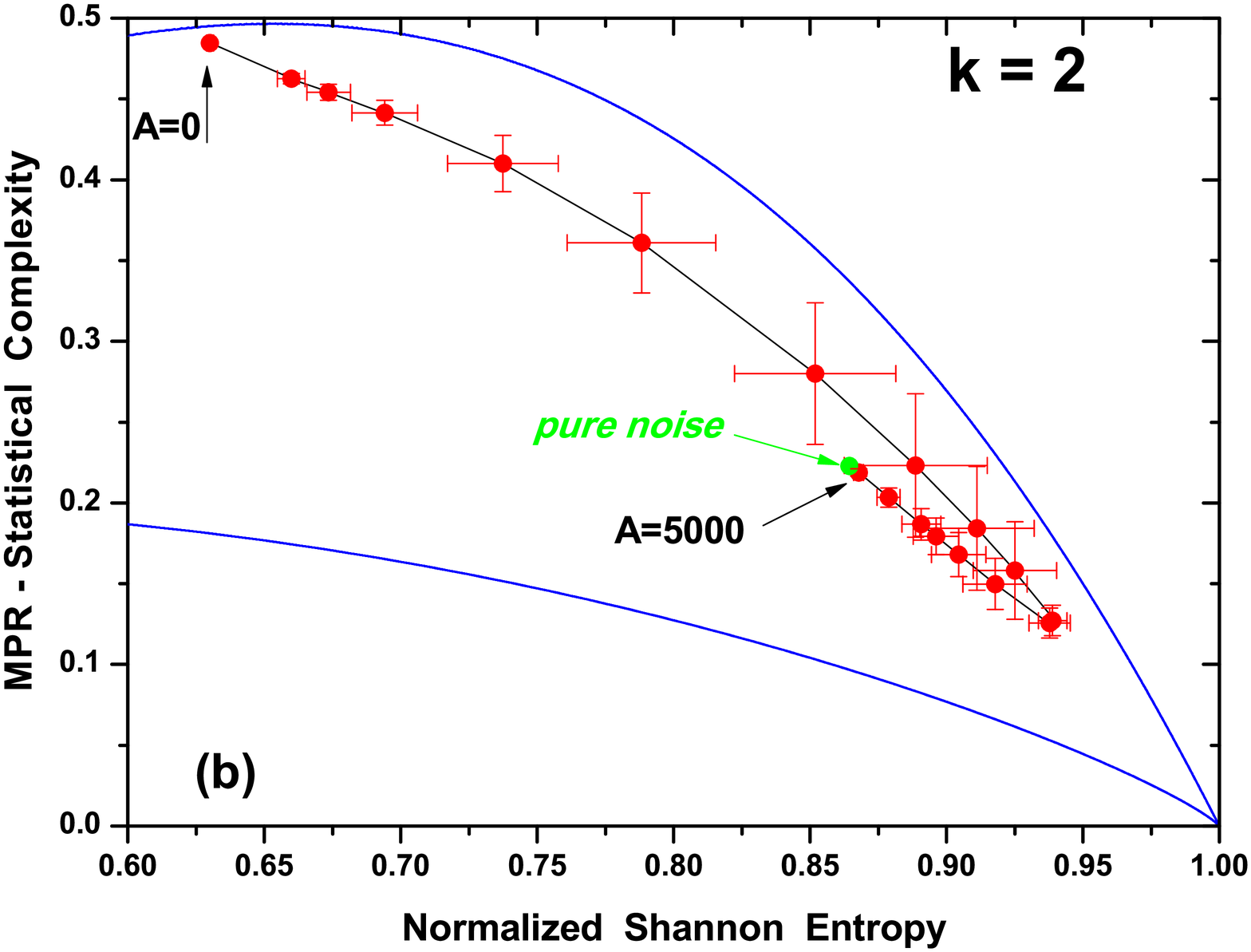}
\caption{
{\it a)\/} The causality entropy-complexity plane for all time series values (mean) corresponding to increasing values of
noise amplitude $A$ and $k=2$.
Value for the case of strictly noisy time series is shown as open symbol.
The  values shown were obtained from times series of length $N=10^5$, 
and Bandt-Pompe parameters $D=6$, $\tau=1$.
The lines represent the values of minimum and maximum statistical complexity ${\mathcal C}_{min}$ 
and ${\mathcal C}_{max}$ evaluated for the case of pattern length $D=6$.
{\it b)\/} Same, showing mean values and standard deviation for each $A$ considered value.
}
\label{plano-k=2}
\end{figure}

\begin{figure}
\noindent
\includegraphics[width=5in]{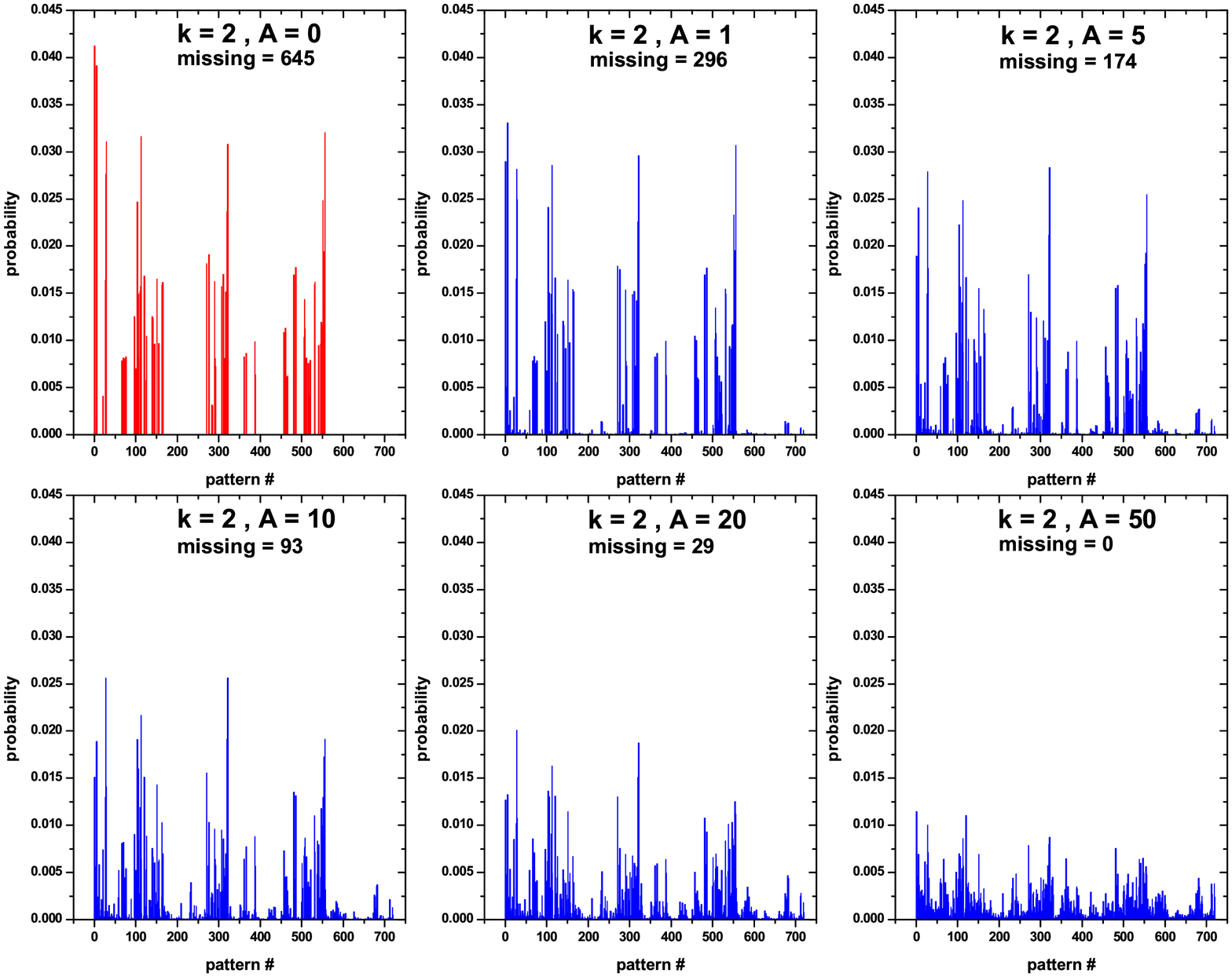}
\includegraphics[width=5in]{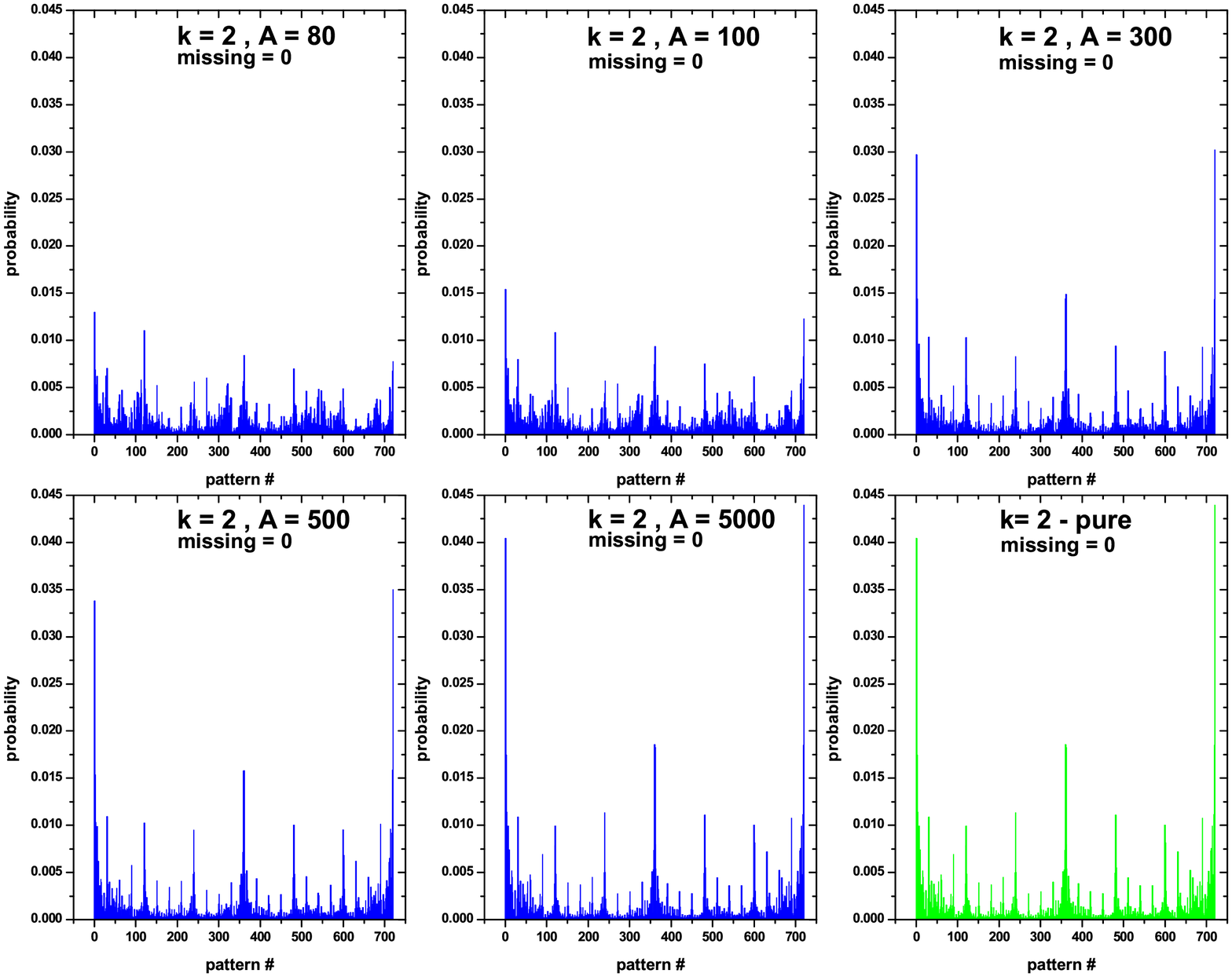}
\caption{
The Bandt-Pompe PDF ($M=720$) for a typical noisy chaotic time series considered (correlated noise $k=2$),
for different increasing values of the noise amplitude $A$.
In the top-left corner the PDF corresponding to the unperturbed logistic map ($A=0$) is presented.
In the bottom-right corner, the PDF for a pure noise $k=2$ time series is display.
The  values shown were obtained from times series of length $N=10^5$, 
and Bandt-Pompe parameters $D=6$, $\tau=1$.
}
\label{PDF-k=2}
\end{figure}

In Figures \ref{plano-k=1} and \ref{plano-k=2} one can appreciate details of the
planar-localization of  noisy chaotic time-series (mean values taken over ten realizations)
that emerge out of the noise-contaminated logistic map. We display values
for  correlated noise with $k=1$ and $k=2$, with increasing noise
amplitude $A \geq 1$.
In these plots we also observe the planar locations of both the time-series for the unperturbed
logistic map ($A=0$) and the pure-correlated noise
($k=1,2$).
As in the previous case of uncorrelated noise ($k=0$), the planar locations move towards the
pure-noise's site (in the present cases, $k=1$ and $k=2$-pure noise, respectively) for increasing
values of the noise-amplitude $A$.
New behaviors emerge, as reflected by  Figures \ref{plano-k=1} and \ref{plano-k=2}.
For varying noise-amplitude values  we appreciate two different trajectories:
1) displacement on the associated planar locations from left to right (starting form the unperturbed
value $A=0$), until reaching a point associated with a critical
value  $A_c$: $A_c \cong 7$ for $k=1$ (see Fig. \ref{plano-k=1}) and $A_c \cong 80$ for $k=2$
(see Fig. \ref{plano-k=2}).
2) At these critical values the trajectory reverses its direction (now it moves from right to left,
``below'' the first curve).
The two trajectories converge at the planar location of the pure-correlated noise.
Identical behavior was observed for all correlated noises with $k$-values in the interval $0<k<2$,
that exhibit specific critical values of $A_c(k)$.

The planar behavior of the noisy, chaotic time-series can be interpreted in terms
of two main interacting effects, namely, those associated with
{\it a)\/} the persistence and robustness of the forbidden patterns in the deterministic chaotic
component of the time-series and,
{\it b)\/} the correlations characterizing  stochastic observational noise (which in the present cases
{\sl do not display \/} forbidden patterns).
These correlations grow with increasing values of $k > 0$.
Looking to the associated  Bandt-Pompe's PDFs behavior, depicted in Figs. \ref{PDF-k=1} and \ref{PDF-k=2}
for different values of $A$, the observed planar  trajectories  can be succinctly described as itemized below.
\begin{itemize}
\item {\it Amplitude range $0 < A \leq A_c$\/}:
We deal with a two scenarios, one (Case A) in which the correlated noise acts as a perturbation
($\mathcal{M}(N,D) \neq 0$).
In the other one (Case B) the logistic and the correlated noise have the same hierarchy
($\mathcal{M}(N,D) = 0$).
Fig. \ref{plano-k=2}.b shows that  for $k=2$ and case A the entropic and complexities' dispersion
(given by the standard deviation) are both small and  due to the perturbative noise.
Contrariwise, for case B the dispersion values increase with the noise amplitude $A$.

\item {\it Amplitude range $A \geq A_c$\/}:
Now the correlated noise dominates. The deterministic component of the time series
can be considered here as a perturbation.
Again,  for this range of noise amplitudes we have ${\mathcal M}(N,D) = 0$.
We also observe that the dispersions of the entropy  ${\mathcal H}_s$ and  the complexity
${\mathcal C}_{JS}$ are small.
\end{itemize}

\section{Conclusions}
\label{sec:Conclusions}

We have revisited in some detail  the paradigmatic concept of
forbidden/missing ordinal patterns and used it as a tool for
distinguishing between deterministic and stochastic behavior in
empiric time-series. In the spirit of
\cite{Amigo2006,Amigo2007,Amigo2008,Zanin2008,Zunino2009,Ouyang2009},
we extended this kind of  analysis by linking it to the physics
described by the causality entropy-complexity plane
\cite{Rosso2007}. Our considerations were made with regards to
deterministic, finite time series contaminated with additive
noises of different degree of correlation. Our analysis of the
noise-determinism competition clearly demonstrates that forbidden
patterns are a deterministic feature of  nonlinear systems and
also reaffirms the usefulness of the entropy-complexity plane as a
powerful tool of the theoretical arsenal.

The planar localization in the causal  entropy-complexity plane
describing an uncontaminated, nonlinear deterministic system is
displaced, by the addition of noise, towards a zone typical of
pure stochasticity. This displacement generates a kind of
trajectory in the ${\mathcal H} \times {\mathcal C}$-plane that is
located in the vicinity of the curve corresponding to maximum
statistical complexity.

If the noise is uncorrelated  (white noise),
the displacement of the system's characteristic point 
in the ${\mathcal H} \times {\mathcal C}$-plane becomes more
pronounced as  the noise intensity (amplitude) increases. It
exhibits a monotonic decreasing behavior (from pure determinism to
pure stochasticity). The noise-effect tends to eliminate forbidden
patterns. Since these, in turn, are features of a deterministic
dynamics, their destruction signals deterministic-nature's loss.

For contaminating correlated noise a new type  of planar
trajectory-behavior emerges as the noise intensity increases.
Starting from a pure deterministic localization, the trajectory
converges to a pure stochastic localization by following a
loop-curve as the noise intensity increases. For a critical value
of noise intensity ($A_c$), however, this trajectory reverses
direction.
One observes a planar-movement  of  the system's characteristic point  
that starts  at the unperturbed value ($A=0$)  and closely
approaches the curve of maximum complexity, from left to right,
with increasing entropic values. At the critical noise intensity
($A_c$) this displacement reverses direction. It takes place now
from right to left, below the original curve and converging to the
planar location typical of pure-correlated noise. The value of the
critical intensity $A_c$ depends on the correlation degree of the
noise and on the deterministic component.

Three different scenarios can be associated:
\begin{enumerate}
\item[{\it a)\/}] The correlated noise acting as a perturbation and ${\mathcal M}(N,D) \neq 0$.
The net noise effect is to destroy the forbidden character of some
of the patterns. However due to the low noise intensity and to its
correlations, the number of affected patterns is relatively low.
The dominance of the deterministic component over the noisy one is
reflected by low dispersion values for both the entropy and the
statistical complexity.
\item[{\it b)\/}] The deterministic and the stochastic components have  the same hierarchy and
${\mathcal M}(N,D) =0$. However, the persistent character
of the forbidden patterns  of the deterministic component and its
interplay with the correlations present
in the noise is reflected in the system's characteristic point 
trajectory, that moves along the curve of maximum complexity,
indicating that  the pertinent patterns do not appear as
frequently as the remaining ones. This behavior is indicative of a
still active deterministic dynamics. The dispersion values for our
two quantifiers increase with the noise intensity. 
\end{enumerate}
The two scenarios above correspond to a noise intensity-range $0 \leq A
\leq A_c$.
\begin{enumerate}
\item[{\it c)\/}]The noisy component is the dominating one and the  deterministic part can
be considered as a perturbation. This scenario corresponds to the
noise intensity range $ A \geq A_c$  and  we have ${\mathcal
M}(N,D)=0$ as well, with low dispersion values for entropy and
statistical complexity.
\end{enumerate}

\section{Appendix}
\label{sec:Appendix}

\subsection{MPR-Statistical Complexity}
\label{sec:Quantifiers1}

Given any arbitrary discrete probability distribution $P = \{ p_i
: i = 1, \cdots ,M \}$, with $M$ the number of freedom-degrees,
Shannon's logarithmic information measure reads \cite{Shannon1949}
\begin{equation}
{\mathrm S}[P] = -\sum_{i=1}^{M}  p_i \ln( p_i) \ .
\label{Shannon}
\end{equation}
The Shannon entropy is a measure of the uncertainty associated to
the physical process described by $P$. It is widely known that an
entropic measure does not quantify the degree of structure or
patterns present in a process \cite{Feldman1998}. Other measures
of statistical or structural complexity, able capture their
organizational properties, are necessary for a better
understanding of chaotic time series \cite{Feldman2008}. Our group
introduced an effective statistical complexity measure (SCM) that
is able to detect essential details of the dynamics and
differentiate different degrees of periodicity and chaos
\cite{Lamberti2004}. This specific SCM, abbreviated as the
MPR-complexity one, provides important additional information
regarding the peculiarities of the underlying probability
distribution, not already detected by the entropy. It is defined,
following the seminal, intuitive notion advanced by L\'opez-Ruiz
{\it et al.\/} \cite{LopezRuiz1995}, via
\begin{equation}
{\mathcal C}_{JS}[P] = {\mathcal Q}_J [P, P_e] \cdot {\mathcal
H}_S[P], \label{Complexity}
\end{equation}
using:
\begin{itemize}
\item [{\it a)\/}] The normalized Shannon entropy ($0 \leq {\mathcal H}_S \leq 1$)
\begin{equation}
{\mathcal H}_S[P] = {\mathrm S}[P] / {\mathrm S}_{max}  \ ,
\label{Shannon-normalizada}
\end{equation}
where ${\mathrm S}_{max} = {\mathrm S}[P_e]  = \ln M$ and $P_e =
\{ 1/M, \cdots , 1/M \}$ the uniform distribution.
\item[{\it b)\/}] The so-called disequilibrium ${\mathcal Q}_J$ ($0 \leq {\mathcal Q}_J \leq 1$),
a quantifier  defined in terms of the extensive
(in the thermodynamical sense) Jensen-Shannon divergence
${\mathcal J} [P, P_e]$ that links two PDFs. We have
\begin{equation}
{\mathcal Q}_J [P, P_e] = Q_0 \cdot {\mathcal J} [P, P_e] \ ,
\label{Disequilibrium}
\end{equation}
with
\begin{equation}
{\mathcal J} [P, P_e] = {\mathrm S} \left[(P + P_e) / 2 \right] -
{\mathrm S}[P] / 2 - {\mathrm S}[P_e] / 2 \ .
\label{Jensen}
\end{equation}
$Q_0$ is  a normalization constant, equal to the inverse of the
maximum possible value of ${\mathcal J} [P, P_e]$. This value is
obtained when one of the values of $P$, say $p_m$, is equal to one
and the remaining $p_i$ values are equal to zero, i.e.,
\begin{equation}
Q_0~=~-2 \left\{ \left( \frac{M+1}{M} \right) \ln(M+1) -2 \ln(2M)
+ \ln M \right\}^{-1} \ .
\label{Q0}
\end{equation}
\end{itemize}

The Jensen-Shannon divergence, that quantifies the difference
between two (or more) probability distributions, is especially
useful to compare the symbol-composition of different sequences
\cite{Grosse2002}. The complexity measure constructed in this way
has the intensive property found in many thermodynamic quantities
\cite{Lamberti2004}. We stress the fact that the statistical
complexity defined above is the product of two normalized
entropies (the Shannon entropy and Jensen-Shannon divergence), but
it is a nontrivial function of the entropy because it depends on
two different probabilities distributions, i.e., the one
corresponding to the state of the system, $P$, and the uniform
distribution, $P_e$.

\subsection{Entropy-Complexity plane}
\label{sec:Plano-HxC} The time-evolution of the MPR-complexity can
be analyzed using a diagram of ${\mathcal C}_{JS}$ versus time
$t$. Thermodynamics' second law states that for isolated systems
the entropy grows monotonically with time ($d{\mathcal H}_S/dt
\geq 0$), entailing  that ${\mathcal H}_S$ can be viewed an
``arrow of time" \cite{Plastino1996}. This suggests to study the
temporal evolution of the SCM  via an analysis of ${\mathcal
C}_{JS}$ versus ${\mathcal H}_S$. A normalized entropy-axis
substitutes for the time-axis. Since one knows  that for a given
value of ${\mathcal H}_S$ the range of possible MPR-complexity
values varies between a minimum ${\mathcal C}_{min}$ and a maximum
${\mathcal C}_{max}$ \cite{Martin2006}, our evolutionary path must
take place in the planar-region delimited by these curves. It was
demonstrated that evolutionary path analysis generates insight
into the details of the system's probability distribution (not
provided by randomness measures like the entropy
\cite{Rosso2007,Feldman2008}) that helps to uncover information
related to the correlational structure between the components of a
given physical process \cite{Rosso2009A,Rosso2009B}.

\subsection{Estimation of the Probability Distribution Function}
\label{sec:PDF-election}

Before attempting the time series characterization using Information Theory quantifiers,
a probability distribution function
(PDF) associated to the time series under analysis should be
provided beforehand. The determination of the most adequate PDF is
a fundamental problem because $P$ and the sample space $\Omega$
are inextricably linked. Many methods have been proposed for a
proper selection of the probability space $(\Omega, P)$. We can
mention: {\it (a)\/} frequency counting \cite{Rosso2009C},
{\it (b)\/} procedures based on amplitude statistics
\cite{DeMicco2008}, {\it (c)\/} binary symbolic dynamics
\cite{Mischaikow1999}, {\it (d)\/} Fourier analysis
\cite{Powell979} and, {\it (e)\/} wavelet transform
\cite{Rosso2001}, among others. Their applicability depends on
particular characteristics of the data, such as stationarity, time
series length, variation of the parameters, level of noise
contamination, etc. In all these cases the dynamics' global
aspects can be somehow captured, but the different approaches are
not equivalent in their ability to discern all the relevant
physical details. One must also acknowledge the fact that the
above techniques are introduced in a rather ``ad hoc fashion" and
they are not directly derived from the dynamical properties
themselves of the system under study.

A better procedure is provided by the symbolic analysis of the time
series, that discretizes the raw  series and transforms it into a
sequence of symbols. Symbolic analysis  is efficient for
nonlinear data processing, with low sensitivity to noise \cite{Finn2003}. Of
course, meaningful symbolic representations of the original series
are not provided by God and require some effort
\cite{Bollt2000,Daw2003}. Today many  people are confident that
the Bandt and Pompe approach for generating PDF's is one of the
most simple symbolization techniques and takes into account
time-causality in the evaluation of the PDF associated to  a time
series \cite{Bandt2002}.
Symbolic data are {\it i)\/} created by associating a
rank to the series-values and {\it ii)\/} defined by reordering
the embedded data in ascending order. Data are reconstructed with
an embedding dimension $D$.
In this way it is possible to quantify
the diversity of the ordering symbols (patterns) derived from a
time series, evaluating the so called ``permutation entropy"
${\mathcal H}_S$, and permutation statistical complexity
${\mathcal C}_{JS}$ (i.e., the normalized Shannon entropy and
MPR-statistical complexity measure evaluated for the Bandt and
Pompe's PDF).

\subsection{The Bandt and Pompe approach for PDF construction}
\label{sec:BP-PDF}

Bandt and Pompe \cite{Bandt2002} introduced a simple and robust method to evaluate the
probability distribution taking into account the time causality of the system dynamics.
They suggested that the symbol sequence should arise naturally from the time series,
without any model-based assumptions.
Thus, they took partitions by comparing the order of neighboring values rather than
partitioning the amplitude into different levels.
That is, given a time series ${\mathcal S} = \{ x_t ; t = 1, \cdots , N \} $, an embedding
dimension $D > 1 ~(D \in {\mathbb N})$,  and an embedding time delay
$\tau$ ($\tau \in {\mathbb N}$), the ordinal pattern of order $D$ generated by
\begin{equation}
\label{eq:vectores}
s \mapsto \left(x_{s-(D-1)\tau},x_{s-(D-2)\tau},\cdots,x_{s-\tau},x_{s}\right) \ ,
\end{equation}
is to be considered.
To each time $s$ we assign a $D$-dimensional vector that results from the evaluation of the
time series at times $s - (D - 1) \tau, \cdots , s - \tau, s$.
Clearly, the higher the value of $D$, the more information about the past is incorporated into
the ensuing vectors.
By the ordinal pattern of order $D$ related to the time $s$ we mean the permutation
$\pi = (r_0, r_1, \cdots , r_{D-1})$ of $(0, 1, \cdots ,D - 1)$ defined by
\begin{equation}
\label{eq:permuta}
x_{s-r_{D-1} \tau} \le  x_{s-r_{D-2} \tau} \le \cdots \le x_{s-r_{1} \tau}\le x_{s-r_0 \tau} \ .
\end{equation}
In this way the vector defined by Eq. (\ref{eq:vectores}) is converted into a unique symbol $\pi$.
In order to get a unique result we consider that $r_i < r_{i-1}$ if $x_{s-r_{i} \tau} = x_{s-r_{i-1} \tau}$.
This is justified if the values of ${x_t}$ have a continuous distribution so that equal values are
very unusual.

For all the $D!$ possible orderings (permutations) $\pi_i$ when the embedding dimension is $D$,
their associated relative frequencies can be naturally computed by the number of times this particular
order sequence is found in the time series divided by the total number of sequences,
\begin{equation}
\label{eq:frequ}
p(\pi_i)~=~ \frac{\sharp \{s|s\leq N-(D-1) \tau ; (s) \quad \texttt{has type}~\pi_i \}}{N-(D-1) \tau} \ .
\end{equation}
In the last expression the symbol $\sharp$ stands for ``number".
Thus, an ordinal pattern probability distribution $P = \{ p(\pi_i), i = 1, \cdots, D! \}$ is obtained
from the time series.

It is clear that this ordinal time-series' analysis entails losing some details of the original
amplitude-information.
Nevertheless,  a meaningful reduction of the complex systems to their basic intrinsic structure
is provided.
Symbolizing time series, on the basis of  a comparison of consecutive points allows for an accurate
empirical reconstruction of the underlying phase-space of chaotic time-series affected by weak
(observational and dynamical) noise \cite{Bandt2002}.
Furthermore, the ordinal-pattern probability distribution is invariant with respect to nonlinear
monotonous transformations.
Thus, nonlinear drifts or scalings artificially introduced by a measurement device do not modify
the quantifiers' estimations, a relevant property for the analysis of experimental data.
These  advantages make the BP approach more convenient than conventional methods based on range partitioning.
Additional advantages of the Bandt and Pompe method reside in its simplicity (we need  few parameters:
the pattern length/embedding dimension $D$ and the embedding time lag $\tau$) and the extremely fast nature
of the pertinent calculation-process \cite{Keller2005,Keller2003}.
We stress that the Bandt and Pompe's methodology is not restricted to time series representative of low
dimensional dynamical systems but can be applied to any type of time series (regular, chaotic, noisy,
or reality based), with a very weak stationary assumption (for $k =D $, the probability
for $x_t < x_{t+k}$ should not depend on $t$ \cite{Bandt2002}).

The probability distribution $P$ is obtained once we fix the embedding dimension $D$ and the embedding
time delay $\tau$.
The former parameter plays an important role for the evaluation of the appropriate probability distribution,
since $D$ determines the number of accessible states, given by $D!$.
Moreover, it has been established that the length $N$ of the time series must satisfy the condition $N \gg D!$
in order to achieve a proper differentiation between stochastic and deterministic dynamics \cite{Rosso2007}.
With respect to the selection of the parameters, Bandt and Pompe suggest in their cornerstone paper
\cite{Bandt2002} to work with $3\leq D \leq 7$ with a time lag $\tau = 1$.
Nevertheless, other values of $\tau$ might provide additional information.
Soriano {\it et al.\/} \cite{Soriano2010a,Soriano2010b} and
Zunino {\it et al.\/} \cite{Zunino2010} recently showed that this parameter is
strongly related, when it is relevant, with the intrinsic time
scales of the system under analysis.
In the present work $D=6$ and $\tau=1$ are used.
Of course it is also assumed that enough data are available for a correct
embedding-time delay procedure (attractor-reconstruction).

\section*{Acknowledgments}
This research has been partially supported by a scholarship from The University of Newcastle awarded to
Laura C. Carpi.
Patricia M. Saco acknowledges support from the Australian Research Council, ARC, Australia.
Osvaldo A. Rosso gratefully acknowledges support from CAPES, PVE fellowship, Brazil.
Mart\'{\i}n G\'omez Ravetti acknowledges support from FAPEMIG and CNPq, Brazil.


\newpage

\newpage

\newpage

\newpage

\newpage

\newpage

\newpage

\end{document}